\documentclass[aps,nofootinbib,twocolumn]{revtex4}
\usepackage{amsfonts, amssymb}
\usepackage{graphicx, epsfig,bm}
\usepackage{color}
\newcommand{\be}{\begin{equation}}
\newcommand{\ee}{\end{equation}}
\newcommand{\bea}{\begin{eqnarray}}
\newcommand{\eea}{\end{eqnarray}}
\newcommand\pp{\,\,\,.}
\newcommand\vv{\,\,\,,}

\begin{document}
\setlength{\unitlength}{1mm}
\author{Gianpiero Mangano}

\affiliation{Istituto Nazionale di Fisica Nucleare - Sezione di
Napoli - \\ Complesso Universitario di Monte S. Angelo, I-80126
Napoli, Italy. }
\title{Shadows of trans-planckian physics on cosmology and the role of the zero-point energy density}

\begin{abstract}
We consider the role of the zero-point energy of a quantum field in cosmology and show that the flow of trans-planckian momenta due to redshift acts as a source for this energy, regularized with a cut-off $\Lambda$ in physical momenta. In order to fulfill Bianchi identity, we generalize Einstein equations, and discuss the corresponding Friedmann homogeneous and isotropic models. In case of a de Sitter phase, such as during inflation, the solution shows a logarithmic behaviour of the Hubble parameter, and a primordial spectrum of scalar perturbations characterized by the spectral index $n_s= 1- \Lambda^2/(3 \pi m_P^2)$ with $m_P$ the Planck mass. We also discuss possible implications of the scenario on  late accelerating stage of the Universe at small redshifts, and the emergence of a fluid characterized by an equation of state $w=P/\rho= -1+ \Lambda^2/(9 \pi m_P^2)$. Primordial perturbation spectrum and dark energy parameter $w$ are thus, predicted to be connected by the simple relation $w=-(2+n_s)/3$. 
\end{abstract}
\bigskip
%\pacs{PACS Numbers: }

\maketitle

\section{Introduction}
Despite of the remarkable success of Einstein theory of gravity applied to the evolution of the Universe, namely the homogenous and isotropic cosmological Friedmann-Robertson-Walker-Lema\^{i}tre (FRWL) model as well as the dynamics of perturbations which give rise to all present observable structures, yet its two extreme energy density regimes in the ultraviolet (UV) and infrared limit are still puzzling. Indeed, a fully satisfactory quantum theory of gravitational interactions at the Planck scale is still lacking. On the other hand, the observation of a recent accelerating phase in the expansion of the Universe from SN-Ia luminosity distance \cite{sn1a1,sn1a2} seems to require that the leading contribution to the energy density budget today is provided by an unknown dark energy component, whose equation of state at low redshifts is very close to that of a cosmological constant, see e.g. \cite{wmap7}. This component has been modelled in quite a large varieties of ways as, just to quote few main examples, scalar field dynamics \cite{quintessence} (as during inflation) or generalization of Brans-Dicke theories, see e.g. \cite{BD} . 

The role of zero-point fluctuation of a quantum field has been usually emphasized as a possible source for the cosmological constant \cite{cosmoconst1,cosmoconst2}, but this introduces a severe hierarchy problem, as its expected value of the order of $\Lambda^4$ with $\Lambda$ a cut-off in $physical$ momenta, is larger by several orders of magnitude than the value of the critical energy today $\rho^0_{cr}$. For $\Lambda \sim $ 1 TeV we have already a too large result, $\Lambda^4 \sim 10^{60} \rho^0_{cr}$. 

In a recent paper \cite{maggiore} it has been pointed out that in analogy with the definition of mass in the Arnowitt-Deser-Misner approach \cite{adm} , the $\Lambda^4$ contribution which is already present in Minkowski spacetime, should be subtracted in order to get the corresponding value in an expanding spacetime. With this procedure the zero-point contribution results to be 
\be
\rho_v= \frac{1}{4 \pi^2} \int_0^{\Lambda a} dy \, y \frac{H^2(t)}{2 a^2} = \frac{H^2(t) \Lambda^2}{16 \pi^2} \vv \label{rhov}
\ee
with $y$ the comoving momentum, $H(t)$ the Hubble parameter and $a$ the scale factor of the FRWL metric, while the corresponding pressure $P_v$ satisfies an equation of state which depends upon the particular phase of the evolution. In fact,  $P_v= -\rho_v/3$ in a de Sitter universe, $P_v=2\rho_v/3 $ during matter domination (MD) and finally, $P_v=\rho_v$ in a radiation dominated (RD) cosmological model. These results hold (at least) for $\rho_v$ representing a sub-dominant component to the total energy density budget, so the equation for modes contributing to $\rho_v$ and $P_v$ can be evaluated on a given background (de Sitter, RD or MD). 

This finding  gives rise to somehow a paradox \footnote{Indeed, the same problem also holds when considering the flat space-time standard term $\rho_v \sim \Lambda^4$, as we discuss below.} when one compares the expected behavior of $\rho_v$ as given by Eq. (\ref{rhov}) with the covariant conservation of the energy momentum tensor 
\be T_{v}^{ab}={\rm  diag} (\rho_v,a^{-2}P_v,a^{-2}P_v, a^{-2} P_v) \pp \ee 
Consider for example the case of  MD Universe. Since $\rho_v \propto H^2$ one gets $\rho_v \sim a^{-3}$. On the other hand
\be 
\nabla_a T_v^{a0} = \dot{\rho_v} + 3 H \left(\rho_v + \frac{2}{3} \rho_v \right) =  \dot{\rho_v} + 5 H \rho_v  \vv
\label{conserv}
\ee
which would rather imply for $\nabla_a T_v^{a0}=0$,  $\rho_v\sim a^{-5}$. A way out to this apparent contradiction is to assume that the zero-point energy density is fueled by a source $J^b$, and $T_v^{ab}$ satisfies the equation,
\be 
\nabla_a T_v^{ab} = J^b \vv \label{conserv2}
\ee
with, during MD, $J^0=2 H \rho_v$, while the spatial component $J^i$ vanish in a homogeneous and isotropic cosmological model. This point has been noticed in \cite{maggiore}, where as a possible solution an ({\it ad hoc}) coupling of zero-point energy-momentum tensor to the other species (matter, radiation) as been invoked, so that still the total $T^{ab}$ is covariantly conserved. We note that the same result for $J^0$ just shown also holds in a de Sitter stage or during RD epoch, so it seems a peculiar property of the zero-point energy density, regardless of the background evolution.

Also in view of this last consideration, we would like to study another possibility, namely that, once we consider the theory regularized with a finite cut-off $\Lambda$ on physical momenta the vacuum energy is sourced by the flow of those modes which, due to the expansion of the Universe, become smaller than $\Lambda$ per unit time or, in a slightly different perspective, to the effect of creation of modes of order of the cut-off  $\Lambda$ at a given time. The {\it rationale} under this approach is that since we can only write our gravity field equations as an effective classical theory which holds for modes with wave-number smaller than $\Lambda$, one should always consider this theory as corresponding to an {\it open} system, and study whether the effect of higher modes might influence the low energy theory when they are redshifted to lower values, smaller than $\Lambda$. 

One might argue that cut-off regularization breaks Lorentz invariance and usually renormalization in curved background is performed with different methods  \cite{birrell,parker,fulling,akh} (see also \cite{maggiore}), whose aim eventually, is to consistently remove divergent terms in the expressions of the energy momentum tensor. On one side, it should be noticed that for the particular case of cosmological models we are interested in, Lorentz invariance is broken anyway by the fact that there is indeed, a preferred (comoving) frame, so that our discussion in the following will be always performed in this frame. On the other hand, we notice that our approach, as we will discuss in the next Sections, leads to predictions which might be testable in the near future, which ultimately is the goal of any physical theory.

\section{The flow of transplanckian modes and effective Einstein equation} \label{sec.2}

Let we illustrate our argument for the standard $\rho_v\sim \Lambda^4$ contribution. In full generality, for any distribution function in phase space $f(P,t)$ describing excitations of a quantum field, the corresponding energy-momentum tensor can be written as (the number $g$ of internal degrees of freedom, such as helicity is set to $g=1$, the generalization to any $g$ being straigthforward)
\be T^a_b = \int \frac{d^4 P}{(2 \pi)^3}  \, \sqrt{-g} f(P,t) \delta ( P^2+m^2) P^a P_b \vv
\ee with the $\delta$ function enforcing mass shell condition, and the four vector $P^a$ is defined as $P^a = m d x^a/ d \lambda$, $x^a$ being the  coordinate and $\lambda$ the affine parameter. They satisfy the condition, $ g_{ab} P^a P^b = - m^2$. If we define the physical momentum and energy as $p^2-E^2 = -m^2$ so that $p^i = a P^i$ in a FRWL spacetime, then integrating over $P^0$ one gets 
\be 
T^a_b = \int \frac{d^3 p}{(2 \pi)^3} f(p,t) \frac{P^a P_b}{P_0} \pp
\ee
Recall that from geodesic motion, physical momentum $p$ redshifts as $1/a$, $P^i$ as $1/a^2$ and the comoving momentum $y_i= g_{ij} P^j$ keeps constant with expansion.
In the particular case of the zero-point energy, $f=1/2$ 
\bea
-T^0_0 &=& \rho_v =\frac{1}{4 \pi^2} \int_0^\Lambda p^2 E(p)  dp \vv  \label{t00} \\
T^i_j &=& \delta^i_j P_v =\delta^i_j  \frac{1}{4 \pi^2}\int_0^\Lambda  p^2 \frac{p^2}{3 E(p) } dp \pp
\eea
By changing the integration variable from $p^i$ to $y^i= a p^i$ we 
have
\bea
-T^0_0 &=&  \rho_v =\frac{1}{4 \pi^2} a^{-4}  \int_0^{\Lambda a} y^2 E(y)  dy \vv \\
T^i_j &=& \delta^i_j P_v = \delta^i_j \frac{1}{4 \pi^2} a^{-4} \int_0^{\Lambda  a} y^2 \frac{y^2}{3 E(y) } dy \vv
\eea
where $E(y) = (y^2 + m^2 a^2 )^{1/2}$. As we are interested in the UV behavior we can neglect in the following the mass $m$ of the field. In this case $P_v= \rho_v/3$. 
The covariant derivative of the energy-momentum tensor, using that $y^i$ is comoving gives for the $0$-component
\be \nabla_a T^{a0} = - 4 H \rho_v + 3 H ( \rho_v + P_v) + H \frac{\Lambda^4}{4 \pi^2}\vv
\ee 
so the first two terms cancel as usual, as $P_v=\rho_v/3$, but to compensate the last term, which comes from time derivative of the upper integration limit and thus, to fulfill conservation of $T^{ab}$, one needs to include a source term $J^0= \Lambda^4 H/(4 \pi^2)$ in the right-hand side of the conservation equation
\be \nabla_a T^{a0} = J^0 \pp \label{cov}
\ee
This source term is provided by modes which become smaller than the cutoff with expansion. The incoming flow is in fact, given 
by the opposite of the energy density of modes with $p=\Lambda$, which leave the trans-Planckian region per unit time
\be  - \frac{1}{4 \pi^2} \Lambda^3 dp = - \frac{1}{4 \pi^2} \Lambda^3 \frac{dp}{dt} dt = H \frac{\Lambda^4}{4 \pi^2} dt \pp \label{sourcestand}
\ee
If we use Eq. (\ref{t00}) we see that the energy density supplied per unit time is  $J^0=4 H \rho_v $. Thus, with a cut-off regularization, we recover the result that the zero-point energy behaves as a cosmological constant, as it is sourced at all times by high modes which become of order $\Lambda$ in the time interval $dt$ and whose contribution exactly compensate the effect of redshift on the sub cut-off modes $p<\Lambda$ in the same $dt$.  Yet, its equation of state is as for radiation, $P= \rho/3$ \cite{branchina}, but there is no contradiction between these two features, as long as one assume that the covariant conservation of energy and momentum is provided by (\ref{cov}). 

Notice that a similar source term would be also present for a thermal bath of excitations with temperature $T$. For $T<\Lambda$ this effect is however exponentially suppressed 
\be 
J^0= \sim  \frac{H}{2 \pi^2} \Lambda^3 E(\Lambda) e^{-E(\Lambda)/T} \pp
\ee
A similar result also holds if we consider $\rho_v$ once the flat space contribution has been subtracted, see Eq. (\ref{rhov}). In this case in fact, we get 
\be J^0 = H \frac{H^2 \Lambda^2}{8 \pi^2} = 2 H \rho_v  \vv
\label{source}
\ee
which is the value we already found in the MD case, see Eq. (\ref{conserv2}) and discussion below it. 

Indeed, as we mentioned already, Eq. (\ref{conserv2}) also gives the correct behavior of $\rho_v$ regardless of the particular expansion phase we consider. To make a further example, during a de Sitter stage $\rho_v \sim$ const and from (\ref{conserv2}) we find 
\be
\dot{\rho_v} + 3 H \rho_v \left(1-\frac{1}{3}\right) =  J^0 = 2 H \rho_v \vv \label{desitter}
\ee
so that $\rho_v=$ const.

The picture which emerges from these considerations is that if we consider the regularized energy momentum tensor with a finite cut-off $T^{ab(\Lambda)}$ the resulting {\it effective} theory does not correspond to an isolated system, as the degrees of freedom larger than $\Lambda$ act as a source due to momentum redshift. This fact enforces a modification of standard Einstein equations, since by Bianchi identities the Einstein tensor $G_{ab}$ satisfies $\nabla_aG^{ab}=0$, while $\nabla_a T^{ab(\Lambda)} \neq 0$. If we keep on defining by $G_{ab}$ the usual  expression, the {\it effective} Einstein equations after cutting modes larger than $\Lambda$ starting from the (unknown) quantum theory of gravity can be written as, by general covariance
\be
G_{ab} + \Sigma_{ab}^{(\Lambda)} = 8 \pi G T_{ab}^{(\Lambda)} \label{Einstein2} \vv
\ee
with $\Sigma_{ab}^{(\Lambda)}$ a 2-tensor satisfying
\be
\nabla_a\Sigma^{ab(\Lambda)}= 8 \pi G \nabla_a T^{ab(\Lambda)} \equiv 8 \pi G J^b \pp
\label{condition}
\ee
In general, $\Sigma_{ab}^{(\Lambda)}$ can be written in terms of the metric, Ricci tensor etc., in a way which of course is not known as we ignore the UV completion of the low energy theory. In the following, we consider the simplest case, very close to the {\it innocent} introduction of a cosmological constant by Einstein himself. In particular, for the homogenous and isotropic models we are interested in, we assume
\be
\Sigma_{ab}^{(\Lambda)} = \sigma^{(\Lambda)}(t) \, g_{ab} \vv
\label{assum}
\ee
where $g_{ab}$ is the standard FRWL metric so that Eq. (\ref{condition}) for $b=0$ gives
\be
-\dot{\sigma}^{(\Lambda)} = 8 \pi G J^0 \pp
\ee
Notice that for the standard case, which as  we already mentioned holds for fluids characterized by thermal excitations such as photons, electrons, etc,, with temperature $T<\Lambda$ we have $J^b\sim0$ and  $\sigma^{(\Lambda)}=$ const., i.e.  the standard Einstein cosmological constant term. 

Using the expression for $J^0$ of Eq. (\ref{source}) (we notice again that we have for the spatial components $J^i=0$ from isotropy), the Friedmann equation now reads
\be
H^2 = \frac{8 \pi G}{3} \left( \rho + \rho_v - 2 \int^a \frac{da'}{a'} \rho_v \right) \vv
\label{friedmann}
\ee
with $\rho$ the standard contribution of matter, radiation and during inflation, the fields driving this expansion stage.

\section{Results from the generalized Friedmann equation}

We start considering a de Sitter Universe, such as during the inflationary stage. If $\rho_v$ represents a sub-dominant contribution, we have from 
(\ref{desitter}) that $\rho_v=$const. Let us consider the role of $\rho_v$ as a small perturbation to $H$. From Eq. (\ref{friedmann})  one gets
\be 
H^2 = \frac{8 \pi G}{3}\left[ \rho_{dS}+ \rho_v\left(1-2 \log \frac{a}{a_0} \right) \right] \vv \label{log}
\ee
with $a_0$ some initial value for $a$ and with $\rho_v << \rho_{dS}$. Thus, the Hubble parameter has a logarithmically decreasing dependence on the scale factor. We define 
\be
\kappa=\frac{G \Lambda^2}{6 \pi} = \frac{\Lambda^2}{6 \pi m_P^2}  \vv
\ee
with $m_P$ the Planck mass. As $\Lambda$ is expected to be at most of the order of $m_P$, this parameter is a small number, $\kappa \sim 0.05$ for $\Lambda= m_P$. Notice that from the expression of $\rho_v$, $\kappa$ also gives the ratio $\rho_{v}/\rho_{dS}$. Using (\ref{rhov}), Eq (\ref{log}) can be rewritten as
\be
H^2 = \frac{8 \pi G}{3} \rho_{dS}\left(1 - 2 \kappa \log\frac{a}{a_0} \right) \vv  \label{log2}
\ee
having rescaled the (arbitrary) initial parameter $\rho_{dS} \rightarrow \rho_{dS}(1- \kappa)$.

It is worth noticing that this result requires a consistency check. Indeed, the equation for the modes $\phi_p$ which contribute to $\rho_v$
\be 
\phi_y''+2 \frac{a'}{a} \phi_y'+ y^2 \phi_y=0 \vv \label{modeeq}
\ee
with the prime denoting derivation with respect to conformal time $\eta$, should be solved self-consistently in the background of both the  $\rho_{dS}$ component and $\rho_v$ itself, while we assumed $\rho_v$ as in (\ref{rhov}), with $H=$ const. In other words, we have to verify that $\rho_v$ behaves as $\sim H^2$ also when the Hubble parameter is slowly varying as in (\ref{log2}), as well as that the source term is given by $J^0=2 H \rho_v$. Only in this case the picture at order $\kappa$ is fully consistent. 

To this end, we start by considering the expression of the scale factor for small $\kappa$
\be 
a(\eta) = -\frac{1+\kappa}{\eta H_0} \left(\frac{\eta}{\eta_0} \right)^{-\kappa} \vv \label{ak}
\ee
with $\eta_0$ some initial value of the conformal time and $H_0=H(\eta_0)$. From this expression we get $a''/a= (2+ 3 \kappa) \eta^{-2}$ and the solution of Eq. (\ref{modeeq}) is
\bea
\phi_y^{(\kappa)}(\eta)&=& \frac{\psi^{(\kappa)}(y \eta)}{a(\eta)} = i \sqrt{\frac{\pi}{2}} \,  \sqrt{y \eta}  \left[\, i J_{3/2 + \kappa}( y \eta) + \right. \nonumber \\ 
&+& \left.   Y_{3/2 + \kappa}( y \eta) \right]\frac{1}{a(\eta)}\vv \label{solution}
\eea
with $J_\lambda$ ($Y_\lambda$) the first (second) kind Bessel function. The normalization has been chosen so that for $\kappa=0$ we have the standard Bunch-Davies vacuum definition for $\eta \rightarrow -\infty$
\be 
\phi_y^{(0)}=\frac{1}{a(\eta)} \left( 1- \frac{1}{y \eta} \right) e^{-i y \eta} \pp \label{BD}
\ee
The well known expression for the energy density
\be \rho_v= \frac{1}{2}  \int_0^{\Lambda a} \frac{d^3 y}{(2 \pi)^3 2 y }\,\left( | \dot{\phi}_y^{(\kappa)} |^2 + \frac{y^2}{a^2} | \phi_y^{(\kappa)} |^2 \right) \vv
\ee
can be worked out as follows. If we define the adimensional variable $s=  y \eta$, we find\be
\rho_v = \frac{1}{8 \pi^2} \left( \frac{\eta}{\eta_0} \right)^{4 \kappa} \frac{H_0^4}{(1+\kappa)^4} \int_0^{S(\eta)} ds \, s^3 \, {\cal P}^{(\kappa)}(s) \vv \label{rhoofs}
\ee
with 
\bea
S(\eta) &=& -\frac{\Lambda}{H_0}(1+\kappa) (\eta_0/\eta)^\kappa \vv \\
{\cal P}^{(\kappa)}(s) & = &(-s)^{-2(1+\kappa)} \nonumber \\
&\times & \left( | d \Psi^{(\kappa)}(s)/ds|^2 + |  \Psi^{(\kappa)} (s)|^2 \right) \vv
\eea
where finally,
\be
\Psi^{(\kappa)} (s) =  \psi^{(\kappa)} (s) (-s)^{1+\kappa} \pp
\ee
For small $\kappa$ Eq. (\ref{rhoofs}) can be evaluated almost entirely in an analitic way, making use of the known expression of $\psi^{(0)}(s)$, see Eq. (\ref{BD}), and at first order in $\kappa$ one obtains the result
\be 
\rho_v = \frac{\Lambda^4}{16 \pi^2} + \frac{\Lambda^2 H^2(\eta)}{16 \pi^2} + \frac{\Lambda^4}{16 \pi^2} {\cal O} (\kappa^2) \vv \label{check}
\ee
with $H(\eta)$ given by (\ref{log2}). The last term correspond to the contribution
\be
 \frac{1}{8 \pi^2} H_0^4 \int_0^{-\Lambda/H_0} ds \, s^3 {\cal P}^{(\kappa)}(s) - \frac{\Lambda^4}{16 \pi^2} \vv
 \ee
 and can be checked numerically that is of order $\kappa^2$.
After subtraction of the $\Lambda^4$ term, we see that the evolution of the vacuum energy density is $\rho_v \sim H^2(\eta)$. It is also quite immediate to check using the definition of $J^0$ given in Eq. (\ref{source}) that still $J^0 = 2 H \rho_v$, and therefore, our scenario seems quite self-consistent. \\

We can now use our  results for $H(a)$ to evaluate the spectrum of scalar perturbations produced during a de Sitter phase, which is usually parameterized as
\be
P(k) k^3 \propto (k/k_0)^{n_s-1} \vv \label{powerlaw}
\ee
with $k_0$ some reference momentum. The amplitude of perturbations with (comoving) momentum $k$ can be estimated as proportional to the value of the squared Hubble parameter at horizon crossing, $k= a H$, which gives
\be
P(k) k^3 \sim P(k_0)\,  k_0^3 \left (1- 2 \kappa \log \frac{k}{k_0} \right)  \pp \label{pk}
\ee
From observations we know that the primordial perturbation spectrum is very close to the flat Harrison-Zeldovich limit. For example, the WMAP Collaboration finds $n_s = 0.963 \pm 0.012$ $(68 \%$ CL) \cite{wmap7}. Expanding (\ref{powerlaw}), from (\ref{pk}) we therefore gets
\be
n_s \sim 1- 2 \kappa = 1- \frac{\Lambda^2}{3 \pi m_P^2} \vv \label{bound}
\ee
which using the result quoted above translates into the result $\Lambda \sim (0.5 \div 0.7) m_P$. 

Notice that if we have in general $N$ massless scalar  fields the value of $\Lambda$ from (\ref{bound}) changes as $\Lambda \rightarrow \Lambda/ \sqrt{N}$ (massless, or very light fermions and gauge bosons do not contribute to $\rho_v$ by conformal invariance, once the flat space contribution has been subtracted). In the most conservative case $N=1$ (the Standard Model Higgs field) and also including the contribution of the two polarization states of the gravitons we find  only a slightly lower results $\Lambda \sim (0.3 \div 0.4) m_P$ . 

Similarly, one can also estimate the running with $k$ of $n_s$, $d n_s/d \log k$ defined as \cite{kos}
\be
P(k) k^3 \propto (k/k_0)^{n_s(k_0)-1+\frac{1}{2}\log(k/k_0) dn_s/d \log k} \vv 
\ee
which is therefore, expected to be of the order of $\kappa^2$ in our model. We recall that observations are fully compatible with a constant $n_s$ \cite{wmap7}.

Summarizing our results, we see that the tilt of primordial perturbations may be the imprint of the role of trans-Planckian mode flow which contribute to the zero-point energy density, while in the standard scenario, it is rather due to the slow rolling down of the inflaton field during the inflationary phase. The fact that $n_s$ is so close to unity, yet significantly different than unity, for $d n_s/d \log k \sim 0$ \cite{wmap7}, would be a signature of the fact that the UV cutoff scale is of the order of Planck mass.\\ 

We now study the behavior of Eq. (\ref{friedmann}) in a RD and MD cosmologies. In these cases the leading term in the Hubble parameter is
\bea 
H &\sim& a^{-2}, \,\,\,\,\,\,\, {\rm (RD)} \vv  \label{RD} \\
H & \sim & a^{-3/2}, \,\,\, {\rm (MD}) \vv \label{MD}
\eea
which are easily recovered in the case $\rho_v$ is very small and substituting its expression, see Eq. (\ref{rhov}) with $H$ as above.
The Friedmann equation can be also rewritten in the form
\be
\left( 1- \kappa \right) H^2 = \frac{8 \pi G}{3} \rho_{R,M} - 2 \kappa \int^a \frac{da'}{a'} H^2 \pp
\ee
Differentiating this equation with respect to $a$, apart from the  behaviours of Eq.s (\ref{RD}) and (\ref{MD}), we notice that it admits also the following solution, which satisfies the corresponding homogeneous equation 
\be
H^2=   {\cal A}\, a^{- 2 \kappa/(1-\kappa)} \vv
\label{result} \ee
with ${\cal A}$ a constant. As long as this contribution is very small, compared to radiation/matter energy density, we thus expect the following behaviours
\bea
H^2 = \frac{8 \pi G}{3} \left( \frac{\rho_{R}}{1-3\kappa /2} +{\cal A} \, a^{- 2 \kappa/(1-\kappa)}\right) \,\, {\rm (RD)} \vv \label{newhubbler} \\
H^2 = \frac{8 \pi G}{3} \left(\frac{\rho_{M}}{1-5 \kappa/ 3} +{\cal A} \, a^{- 2 \kappa/(1-\kappa)}\right) \,\, {\rm (MD)} \pp \label{newhubblem} 
\eea

The expansion is thus driven by two fluids, an ordinary radiation/matter term, with an energy density which can be adjusted to match the measured value $\Omega_M$ and $\Omega_R$ by rescaling  $\rho_R \rightarrow \rho_R(1-3 \kappa/2)$ and  $\rho_M \rightarrow \rho_M(1-5 \kappa/3)$ and a new contribution whose effective equation of state parameter $w=P/\rho$ in terms of $\kappa$  is given by
\be
w = -\frac{1-5 \kappa/3}{1-\kappa} \simeq  -1 + \Lambda^2/(9 \pi m_P^2) \pp \label{weff}
\ee
Since $\kappa$ is small, we see that this new term plays the role of a  cosmological (almost) constant. Indeed, for $\Lambda= m_P$ we have $\kappa \sim 0.05$ and  $w \sim -0.96$.

The result of (\ref{weff}) is quite intriguing, showing the interplay of UV physics, represented by the cut-off scale $\Lambda$, and  the late and much lower energy regime of the expansion of the Universe, characterized by the emerging role of a tiny cosmological constant. Again, as in the case of scalar perturbation spectrum in de Sitter phase, the fact that $w$ is so close to unity is due to that fact that $\Lambda \leq m_P$. 

Notice that since for small ${\cal A}$ and decreseasing $a$ the last term  is increasingly smaller than radiation and matter,  we expect that the radiation-matter equality redshift would be basically unchanged, as well as the whole dynamics of structure formation and Cosmic Microwave background anisotropy peaks. Similarly, during RD there are no particular new features in the expansion behavior, thus we have no bounds on 
$\kappa$ from, say, Big Bang Nucleosynthesis \cite{iocco}. 

It would be tempting to interpret the late time behavior of Eq.  (\ref{newhubblem}) as a possible solution for the present accelerating phase of the Universe expansion, by choosing the constant ${\cal A}$ of the correct order of magnitude such to reproduce the observed value of $\Omega_\Lambda\sim 0.75 $ today. Unfortunately, since ${\cal A}$ should be tuned  as ${\cal A} \sim \rho^0_{cr}$, we cannot presently give a simple explanation of the {\it coincidence} problem. Notice that when  $\rho_v$  starts dominating the energy density, the expansion behaves as in an (almost) de Sitter model. From our previous analysis, see Eq. (\ref{check}), it follows that the behaviour $H \sim a^{-2 \kappa}$ is a consistent solution, since $\rho_v \propto H^2$, as well as $J^0 = 2 H \rho_v$.

\section{Conclusions}

In this paper we have considered some possible cosmological effects of the zero-point energy density $\rho_v$. Regularizing its contribution with a momentum cut-off $\Lambda$, we have emphasized the role of flow of trans-planckian modes which become smaller than $\Lambda$ per unit time due to redshift in order to obtain a consistent time behaviour of $\rho_v$. They act as an energy density flow for an effective  \'a la Wilson low energy gravity theory, which calls for a modification of Einstein equations in order to fulfill Bianchi identities. 

In a homogenous and isotropic Universe we have considered the corresponding Friedmann equation and studied the Hubble parameter behaviour when the effect of $\rho_v$ is considered at first order in the parameter $\kappa = \Lambda^2/(6 \pi m_P^2)$. The value of $\kappa$ is expected to be small as long as the cut-off scale is smaller or of the order of  the Planck mass scale, where indeed large quantum gravity effects are expected to be quite relevant. 

Interestingly, the UV parameter $\kappa$ seems to leave its imprint in the Universe evolution. In particular, we have pointed out how the spectral index $n_s$ of primordial perturbation produced during the inflationary epoch may be simply related to $\kappa$ as $n_s = 1-2 \kappa$, and how the late accelerating phase of the Universe for small redshifts $z\leq 1$ could be explained in terms of the contribution of $\rho_v$. The latter gives rise to an energy density contribution characterized by an equation of state $w=P/\rho=-1+2 \kappa/3 = -1+\Lambda^2/(9 m_P^2)$, showing an intriguing interplay of UV physics with the low energy behaviour of the Universe evolution. These two results can be combined in a relationship between these two parameters, namely $w= -(2+n_s)/3$, which in the standard case are in general, unrelated.

\begin{acknowledgments}

This work is supported by the {\it Istituto Nazionale di Fisica Nucleare} I.S. FA51 and the PRIN 2010 ``Fisica
Astroparticellare: Neutrini ed Universo Primordiale'' of the Italian {\it Ministero dell'Istruzione, Universit\`a e Ricerca}. I am delighted to warmly thanks V. Branchina for bringing my interest to the vacuum energy density problem and for discussions, M. Maggiore for inspiring comments, as well as the {\it Astroparticle Group} at University of Naples for discussions. 

\end{acknowledgments}


\begin{thebibliography}{99} 
\bibitem{sn1a1} S. Perlmutter {\it et al.} (Supernova Cosmology Project), Ap.J. {\bf 517},  565 (1999), [arXiv:astro-ph/9812133].
\bibitem{sn1a2} A.G. Riess  {\it et al.} (Supernova Search Team), Ap.J. {\bf 659}, 98 (2007), [arXiv:astro-ph/0611572].
\bibitem{wmap7} E.~Komatsu {\it et al.} (WMAP), arXiv:1001.4538 [astro-ph.CO].
\bibitem{quintessence} P.J.E. Peebles and B. Ratra, Rev. Mod. Phys. {\bf 75}, 559 (2003), [arXiv:astro-ph/0207347].
\bibitem{BD} F. Perrotta, C. Baccigalupi and S. Matarrese, Phys. Rev.  D {\bf 61}, 023507 (2000), [arXiv:astro-ph/9906066].
\bibitem{cosmoconst1} S. Weinberg, Rev. Mod. Phys. {\bf 61}, 1 (1989).
\bibitem{cosmoconst2} T. Padmanabhan, Phys. Rept. {\bf 380}, 235 (2003), [arXiv:hep-th/0212290].
\bibitem{maggiore} M. Maggiore, arXiv:1004.1782 [astro-ph].
\bibitem{adm} R.L. Arnowitt, S. Deser and C.W. Misner (1962). In {\it Gravitation: an introduction to current research}, L.Witten ed., Wiley, NY, [arXiv:1004.1782].
\bibitem{birrell} N.D. Birrell and P.C.W. Davies, {\it Quantum fields in curved space}, Cambridge University Press, (1982).
\bibitem{parker} L. Parker and S. Fulling, Phys. Rev. D {\bf 9}, 341 (1974).
\bibitem{fulling} S.A. Fulling, L. Parker and B.L. Hu, Phys. Rev. D {\bf 10}, 3905 (1974).
\bibitem{akh} E.K. Akhmedov, arXiv:hep-th/0204048.
\bibitem{branchina} V. Branchina and D. Zappala, arXiv:0705.2299[hep-ph].
\bibitem{kos} A. Kosowsky and M.S. Turner, Phys Rev D {\bf 52}, 1739 (1995).
\bibitem{iocco} F. Iocco, G. Mangano, G. Miele, O. Pisanti and P.D. Serpico, Phys.  Rept.  {\bf 472}, 1 (2009), [arXiv:0809.0631 [astro-ph].
\end{thebibliography}
\end{document}